\documentstyle[aps,prc,epsfig]{revtex}

\begin{document}

\def\beq{\begin{equation}}
\def\eeq{\end{equation}}
\def\beqa{\begin{eqnarray}}
\def\eeqa{\end{eqnarray}}
\def\MeV{\nobreak\,\mbox{MeV}}
\def\GeV{\nobreak\,\mbox{GeV}}
\def\keV{\nobreak\,\mbox{keV}}
\def\fm{\nobreak\,\mbox{fm}}
\def\Tr{\mbox{ Tr }}
\def\La{\Lambda}
\def\ka{\kappa}
\def\la{\lambda}
\def\ga{\gamma}
\def\Ga{\Gamma}
\def\om{\omega}
\def\rh{\rho}
\def\si{\sigma}
\def\ps{\psi}
\def\ph{\phi}
\def\de{\delta}
\def\al{\alpha}\def\be{\beta}
\def\lb{\label}
\def\nn{\nonumber}
\newcommand{\rag}{\rangle}
\newcommand{\lag}{\langle}
\newcommand{\bph}{\mbox{\bf $\phi$}}
\newcommand{\rf}{\ref}
\newcommand{\ct}{\cite}
\newcommand{\ii }{\'{\i}}
\def\me#1{\langle{#1}\rangle}
\def\meadjust#1{\left<{#1}\right>}
\def\bra#1{\langle #1|}
\def\ket#1{| #1\rangle}
\def\qbar{\overline{q}}
\def\gs{g_{\rm s}}
\def\G{{\cal G}}
\def\mixbar{\gs\qbar\sigma\!\cdot\!\G q}
\def\mixs{\gs\bar{s}\sigma\!\cdot\!\G s}
\def\alphas{\alpha_{\rm s}}
\def\gluoncon{{\displaystyle{\gs^2G^2}}}
\def\qslash{\rlap{/}{q}}
\def\xsla{\rlap{/}{x}}
\def\qsq{q^2}
\def\pli{p^\prime}
\def\mli{{M^\prime}^2}

\title{Hadronic form factors and the $J/\psi$ secondary production 
cross section: an update}
\author{$^1$F. Carvalho, $^2$F.O. Dur\~aes, $^3$F.S. Navarra and $^3$M. Nielsen}
\address{$^1$Instituto de F\ii sica Te\'orica, Universidade Estadual Paulista\\
Rua Pamplona, 145, 01405-000, S\~ao Paulo - S.P.\\
$^2$Universidade Presbiteriana Mackenzie, C.P. 01302-907 S˜ao Paulo, Brazil
$^3$Instituto de F\'{\i}sica, Universidade de S\~{a}o Paulo\\
 C.P. 66318,  05315-970 S\~{a}o Paulo, SP, Brazil}
\maketitle

\begin{abstract}
Improving previous calculations, we compute the 
$D + \bar{D} \rightarrow  J/\psi + \pi $ cross section using the most complete 
effective lagrangians available. The new crucial ingredients are the form factors on the 
charm meson vertices, which are determined from QCD sum rules calculations. Some 
of them became available only very recently and the last one, needed for our present 
purpose, is calculated in this work.  
\end{abstract} 


\vspace{1cm}
\section{Introduction}

Before RHIC, RHIC physics was relatively simple. We were basically searching for 
a quasi-ideal gas of deconfined quarks and gluons(QGP). One of the best signatures of 
this new state of matter was charmonium suppression  \cite{ma86}. During the 
last four years, due to intense work both theoretical and experimental, this naive 
picture changed drastically. On the theoretical side, careful numerical simulations 
\cite{thews,rapp,polleri,brat} have shown that, due to the importance of charm 
recombination in the 
deconfined phase
and also to  final state interactions, the number of $J/\psi$'s may stay approximately 
the same. From the experimental side, especially from the analysis of elliptic flow, 
came the conclusion that the new state of matter is not a gas, being rather a strongly 
interacting fluid, now called sQGP \cite{shu}.   Taking the existing 
calculations seriously,  
it is no longer clear that an overall suppression of the number of $J/\psi$'s 
will be a signature of QGP.  
A more complex pattern can emerge, with suppression in some regions of the 
phase space and enhancement in others \cite{huf,dnn}. Whatever the new QGP 
signature (involving charm) turns out to be, it is necessary to understand 
better the mechanisms of  $J/\psi$ production and dissociation  by 
collisions with comoving hadrons. 

A great effort has been dedicated to understand  $J/\psi$ dissociation in
a hadronic environment. Since there is no direct experimental information on 
$J/\psi$ absorption cross sections by hadrons, several theoretical approaches have 
been proposed to estimate their values.
One approach was based on charm quark-antiquark dipoles interacting with the 
gluons of a larger (hadron target) dipole. This is the  Bhanot-Peskin (BP) approach 
\cite{bhp}, which was rediscovered by Kharzeev and Satz \cite{kha2} in the 
mid-nineties and updated \cite{lo,arleo} in the last years.  Finally, 
the recent next to leading order calculations presented in \cite{song} have conclusively
shown that, for charmonium, the formalism breaks down because this system is not heavy 
enough. Also considered was quark exchange between two (hadronic) bags 
\cite{wongs,mbq}. The most explored approach has 
been the meson exchange mechanism  \cite{mamu98,osl,haglin,linko,nnr}. In our 
opinion the most reliable calculations of $ \sigma_{J/\psi -  \pi} $
 was done with QCD sum rules \cite{qcdsr}. However, due to a low momentum approximation, the validity of
this calculation was restricted  to low energy reactions, close to the dissociation 
threshold. This is probably not enough for the numerical simulations mentioned above. Therefore, 
in order to have cross sections valid at higher energies, the effective lagrangrian approach still
remains the most appropriate option.

After many works on the subject some consensus has been achieved, 
%``...the state of the theory of interactions  topic''. 
at least in what concerns the determination of the order of magnitude, which, 
in the case of the $J/\psi$ pion interaction, is determined to be 
$ 1 \, < \,  \sigma_{J/\psi -  \pi} \, < \, 10 $ mb in the energy region close 
to the open charm production threshold.

Once the $J/\psi$ dissociation cross section is known, using detailed balance one
can attempt to estimate the charmonium formation cross section through the fusion 
of open charm, as, for example, $D \bar{D}\rightarrow J/\psi+\pi$.  This is known as
secondary charmonium production. As it was first pointed out in \cite{ko98}, in 
nucleus - nucleus collisions at sufficiently high 
energies, the number of produced $D$ and $D^*$ mesons increases and also 
the lifetime of the hadronic fireball increases. It becomes then possible that a 
significant number of $J/\psi$'s be formed by open charm fusion.  Later, an estimate 
made in \cite{redlich} indicated that this mechanism would be relevant only for
LHC energies. The authors stressed, however, that their conclusion was very sensitive 
to the value of the $J/\psi$ formation cross section, or equivalently to the 
absorption cross section, which in that case was the one computed with the 
Bhanot-Peskin approach. The subject was left aside for some time. Recently, after the
revision of the $J/\psi$ absorption cross section to larger values, secondary 
$J/\psi$ production was incorporated in event generators in \cite{zhang2002} and 
\cite{brat}. According to these simulations, the number of secondary $J/\psi$'s is 
significant already at RHIC energies.

Given the renewed interest on the subject we shall, in this work, further refine our
estimate of the $J/\psi$ interaction cross section, giving now emphasis to secondary 
charmonium production. We shall employ effective lagrangians with form factors 
calculated with  QCDSR. In particular, in the present calculation we shall make use of 
the $D^* D^* J/\psi$ form factor, which was obtained only very recently \cite{miru} 
and we shall also calculate the $D^* D^* \pi$ form factor, which had not been 
calculated so far.

\section{The effective lagrangians}

Since the pioneering work of Muller and Matynian \cite{mamu98}, there has been an 
intense discussion concerning the details and properties of the effective 
lagrangians which describe the interactions among charm mesons. Here we do not add 
anything new to this discussion. We shall use what we believe is the 
state - of - the - art lagrangian. For the sake of completeness and for future use 
we present  below the effective lagrangians considered in this work:
\beqa
{\cal L}_{D^*D\pi}&=&ig_{D^*D\pi}(D_{\mu}^*\partial^{\mu}\pi\bar{D}-
D\partial^{\mu}\pi\bar{D^*_{\mu}})
\label{dsdpi}\\
{\cal L}_{\psi D^{*}D}&=&g_{\psi D^{*}D}\epsilon^{\mu \nu \alpha \beta}
\partial_{\mu}\psi_{\nu}\left(\partial_{\alpha}{D}^{*}_{\beta}\bar{D}+D\partial_{
\alpha}\bar{D^{*}}_{\beta}\right)
\label{psidsd}\\
{\cal L}_{\psi D D\pi}&=&ig_{\psi D D \pi}\epsilon^{\mu \nu \alpha \beta}
\psi_{\mu}\partial_{\nu}D\partial_{\alpha}{\pi}\partial_{\beta}\bar{D}
\label{psiddpi}
\eeqa

\begin{eqnarray}
{\cal L}_{\psi DD}&=&-ig_{\psi DD}~ \psi^\mu 
\left ( \partial_\mu D \bar{D}  - D\partial_\mu \bar{D} \right )
\label{psidd}\\
{\cal L}_{D^*D^*\pi}&=&-g_{D^*D^* \pi}\epsilon^{\mu \nu \alpha \beta}
\partial_{\mu}D^*_{\nu}{\pi}\partial_{\alpha}\bar{D^*}_{\beta}
\label{dsdspi}\\
{\cal L}_{\psi D^*D^*}&=& ig_{\psi D^*D^*}~
\left [ \psi^\mu \left ( \partial_\mu D^{* \nu} \bar {D^*_\nu} 
- D^{* \nu} \partial_\mu \bar {D^*_\nu} \right )\right.\nonumber \\
&+&\left ( \partial_\mu \psi_\nu D^{*\nu} -\psi_\nu 
\partial_\mu D^*_\nu 
\right )\bar {D^{* \mu}}
+\left. D^{* \mu} \left ( \psi^\nu \partial_\mu \bar {D^*_\nu} -
\partial_\mu \psi_\nu \bar {D^{*\nu}} \right ) \right ]
\label{psidsds}\\
{\cal L}_{\psi D^* D\pi}&=&-g_{\psi D^* D \pi}
\psi^{\mu}\left(D\pi\bar{D^*}_{\mu}+D^*_{\mu}\pi\bar{D}\right)
\label{psidsdpi}
\eeqa

\beqa
{\cal L}_{\psi D^* D^*\pi}=ig_{\psi D^* D^* \pi}\epsilon^{\mu \nu \alpha \beta}
\psi_{\mu}{D^*}_{\nu}\partial_{\alpha}\pi\bar{D^*}_{\beta}+ih_{\psi D^* D^* \pi}
\epsilon^{\mu \nu \alpha \beta}\partial_\mu \psi_\nu D^{*}_{\alpha}\pi{D^*}_{\beta}
\eeqa

With these lagrangians were are able to compute the processes
$D \bar{D}\rightarrow J/\psi+\pi$, which involves the diagrams in
Figure 1, the process $D^* \bar{D}\,\rightarrow J/\psi + \pi$,   
corresponding to the diagrams shown in Figure 2  and also the process
$D^* \bar{D^*}\,\rightarrow J/\psi + \pi$, corresponding to the diagrams 
in Figure 3.

Calling $p_1$ and  $p_2$ the four-momenta of the incoming particles and 
$p_3$ and  $p_4$  the four-momenta of the outgoing particles, we can derive the 
Feynman rules from the above lagrangians and obtain the invariant amplitudes for each 
one of the processes in Figures 1, 2 and 3. They are given by:
\beqa
{\cal M}^{1a}_{\mu}=-g_{D^* D \pi}\,p_{3_{\alpha}}\,
\frac{1}{t-m^2_{D^*}}\left(g_{\alpha\beta}-\frac{q_{\alpha}q_{\beta}}{m_{D^*}}\right)\,
g_{\psi D^* D }\,\epsilon^{\rho \mu \theta \beta}\,q_{\theta}\,p_{4_{\rho}}
\label{m1a}
\eeqa
\beqa
{\cal M}^{1b}_{\mu}&=&g_{D^* D \pi}\,p_{3_{\alpha}}\,
\frac{1}{u-m^2_{D^*}}\left(g_{\alpha\beta}-\frac{(p_2-p_3)_{\alpha}(p_2-p_3)_{\beta}}
{m_{D^*}}\right)\,\,\,\times \nonumber\\
&&\,g_{\psi D^* D }\,\epsilon^{\rho \mu \theta \beta}\,(p_2-p_3)_{\theta}\,p_{4_{\rho}}
\label{m1b}
\eeqa
\beqa
{\cal M}^{1c}_{\mu}=g_{\psi D^* D \pi}\,
\epsilon^{\mu \rho \theta \delta}\,p_{1_{\rho}}\,p_{2_{\delta}}\,p_{3_{\theta}}
\label{m1c}
\eeqa

\beqa
{\cal M}^{2a}_{\nu \sigma}=g_{D^* D \pi}\,p_{3_{\nu}}\,
\frac{1}{t-m^2_{D}}\,
\,g_{\psi D D }\,(p_2-q)_{\sigma}
\label{m2a}
\eeqa
\beqa
{\cal M}^{2b}_{\nu \sigma}=g_{\pi D^* D^*}\,\epsilon^{\gamma \nu \delta\alpha} 
p_{1_{\gamma}}\,q_{\delta}\,
\frac{1}{t-m^2_{D^*}}\left(g_{\alpha\beta}-\frac{q_{\alpha}q_{\beta}}{m_{D^*}}\right)\,
\,g_{\psi D^* D }\,\epsilon^{\rho \sigma \theta \beta}p_{4_{\rho}}\,q_{\theta}
\label{m2b}
\eeqa
\beqa
{\cal M}^{2c}_{\nu \sigma}=g_{D^* D \pi}\,p_{3_{\alpha}}\,
\frac{1}{u-m^2_{D^*}}\left(g_{\alpha\beta}-\frac{(p_2-p_3)_{\alpha}(p_2-p_3)_{\beta}}
{m_{D^*}}\right)\,\,\,\times \nonumber\\
\,g_{\psi D^* D^*}\,\bigg[(p_1-p_2+p_3)_{\sigma}\,g_{\nu \beta}
\,-\,(p_1+p_4)_{\beta}\,g_{\nu \sigma}\,+\,(p_2-p_3+p_4)_{\nu}\,g_{\sigma \beta}\bigg]
\label{m2c}
\eeqa
\beqa
{\cal M}^{2d}_{\nu \sigma}=-g_{\psi D^* D \pi}\,g_{\nu \sigma}
\label{m2d}
\eeqa

\beqa
{\cal M}^{3a}_{\mu \nu \sigma}=g_{D^* D \pi}\,p_{3_{\mu}}\,
\frac{1}{t-m^2_{D}}\,
\,g_{\psi D^* D }\,\epsilon^{\rho \sigma \theta \nu}\,p_{4_{\rho}}\,p_{2_{\theta}}
\label{m3a}
\eeqa
\beqa
{\cal M}^{3b}_{\mu \nu \sigma}=-g_{D^* D \pi}\,p_{3_{\nu}}\,
\frac{1}{u-m^2_{D}}\,
\,g_{\psi D^* D }\,\epsilon^{\rho \sigma \theta \mu}\,p_{4_{\rho}}\,p_{1_{\theta}}
\label{m3b}
\eeqa
\beqa
{\cal M}^{3c}_{\mu \nu \sigma}=g_{\pi D^* D^*}\,\epsilon^{\delta \mu \theta \alpha} 
p_{1_{\delta}}\,q_{\theta}\,
\frac{1}{t-m^2_{D^*}}\left(g_{\alpha\beta}-\frac{q_{\alpha}q_{\beta}}{m_{D^*}}\right)
\,\,\,\times
\nonumber\\
\,g_{\psi D^* D^*}\,\bigg[(q-p_2)_{\sigma}\,g_{\nu \beta}
\,+\,(p_2+p_4)_{\beta}\,g_{\nu \sigma}\,-\,(p_4+q)_{\nu}\,g_{\sigma \beta}\bigg]
\label{m3c}
\eeqa
\beqa
{\cal M}^{3d}_{\mu \nu \sigma}=-g_{\pi D^* D^*}\,\epsilon^{\delta \nu \theta \alpha} 
p_{2_{\theta}}\,(p_2-p_3)_{\delta}\,
\frac{1}{u-m^2_{D^*}}\left(g_{\alpha\beta}-\frac{(p_2-p_3)_{\alpha}(p_2-p_3)_{\beta}}
{m_{D^*}}\right)\,\,\,\times \nonumber\\
g_{\psi D^* D^*}\,\bigg[(p_2-p_3-p_1)_{\sigma}\,g_{\mu \beta}
\,+\,(p_1+p_4)_{\beta}\,g_{\mu \sigma}\,-\,(p_2-p_3+p_4)_{\mu}\,g_{\sigma \beta}\bigg]
\label{m3d}
\eeqa
\beqa
{\cal M}^{3e}_{\mu \nu \sigma}=-g_{\psi D^* D^* \pi}\,\epsilon^{\sigma\mu \theta \nu} 
p_{3_{\theta}}\,+\,h_{\psi D^* D^* \pi}\,\epsilon^{\sigma\mu \theta \nu} p_{4_{\theta}}
\label{m3e}
\eeqa

Finally the cross section for these processes is obtained with:
\beqa
\frac{d\sigma}{dt}=\frac{1}{N}\frac{1}{64 \pi \,{\bf p}^2_i}\,
\sum_{spin}\left|{\cal M}^2\right|
\eeqa
where ${\bf p}^2_i$ a three-momentum squared in the center of mass system  
and the factor $\displaystyle \frac{1}{N}$ comes from the average over the initial 
state polarizations.

As extensively discussed in previous works, although the above lagrangians and 
amplitudes are quite satisfactory from the point of view of symmetry requirements, 
their straightforward application to the computation of cross sections leads to 
unacceptably large results. This comes from the fact that the exchanged particles may 
be far off-shell and therefore they enter (or leave) a vertex with a very  different 
resolving power. In one extreme case  considered in the recent past 
\cite{psidd}, a virtual 
$J/\psi$ probing a $D$ meson, had the behavior of a parton (!). Of course, when this 
happens, the compact $J/\psi$ almost misses the large $D$ and  as a consequence the
cross section of the whole process drops significantly. This physics of  spatial 
extension and resolving power is contained in the form factors. It has been realized by 
many authors that calculations with and without form factors lead to results differing 
by up to two orders of magnitude! Therefore we simply {\it can not ignore the form 
factors}. We must include them in order to obtain reliable results! 

Looking at the diagrams in Figures 1, 2 and 3 we notice that we need the following 
form factors (and the corresponding coupling constants):
\beq
g^{(D^{*})}_{\pi D D^{*}}(t) \,\,\,\,\,\,
g^{(D)}_{\psi D D}(t) \,\,\,\,\,\,
g^{(D^{*})}_{\psi D D^{*}}(t)  \,\,\,\,\,
g^{(D)}_{\psi DD^{*}}(t)  \,\,\,\,\,
g^{(D^{*})}_{\psi D^{*} D^{*}}(t)  \,\,\,\,\,
g^{(D^{*})}_{\pi D^{*} D^{*}}(t) 
\label{list}
\eeq
where $t$ is the usual momentum transfer squared and in the superscript in 
parenthesis we denote the off-shell particle. This is an important distinction, 
because the form factors in the same vertex are very different when different 
particles  are off-shell. The most reliable way to compute these factors is with 
the use of the QCD sum rules techniques. The first one of the list was calculated 
in \cite{dsdpi}, the second in \cite{psidd}, the third and fourth in \cite{dsdpsi},  
the fifth in \cite{miru} and they  read:
\beqa
g^{(D)}_{\pi D^*D}(t)=17.9\left(\frac{(3.5 \,GeV)^2-m_D^2}{(3.5 \,GeV)^2-t}
\right)=h_4(t,m_D^2)
\label{h1}
\eeqa
\beqa
g^{(D)}_{\psi DD}(t)=5.8
\left(e^{-\left(\frac{20-t}{15.8}\right)^2}\right)=h_3(t)
\label{h2}
\eeqa
\beqa
g^{(D^{*})}_{\psi DD^{*}}(t)=20
\left(e^{-\left(\frac{27-t}{18.6}\right)^2}\right)=h_1(t)
\label{h3}
\eeqa
\beqa
g^{(D)}_{\psi DD^{*}}(t)=13\left(e^{-\left(\frac{26-t}{21.2}\right)^2}\right)=h_2(t)
\label{h4}
\eeqa
\beqa
g^{(\pi)}_{\pi D^*D^*}(t)=4.8\left(e^{\left(\frac{t}{6.8}\right)}\right)=
h_6(t,m_{\pi}^2)
\label{h5}
\eeqa
\beqa
g^{(D^*)}_{\psi D^*D^*}(t)=6.2\left(e^{\left(\frac{t}{3.55}\right)}\right)=h_5(t)
\label{h6}
\eeqa

The last form factor in (\ref{list}) will be calculated below.

\section{The $\pi D^{*} D^{*}$ form factor}

In this section we shall, for the first time, compute the $\pi D^{*} D^{*}$ form 
factor using QCDSR \cite{svz,rry}.  
In this approach, the short range perturbative QCD is
extended by an OPE expansion of the correlators, which results in 
a series in powers of
the squared momentum with Wilson coefficients. The convergence at low
momentum is improved by using a Borel transform. The expansion involves
universal quark and gluon condensates. The quark-based calculation of
a given correlator is equated to the same correlator, calculated using
hadronic degrees of freedom via a dispersion relation, providing sum rules
from which a hadronic quantity can be estimated. 

We shall use the three-point function  to evaluate the 
$D^*D^*\pi$  form factor for an off-shell $D^*$ meson, following the
procedure suggested in ref.~\cite{rhodd} and further extended in
\cite{dsdpsi}. This means that we shall calculate the correlators 
for a $D^*$ off-shell and then for a $\pi$ off-shell, requiring that 
the corresponding extrapolations to the respective poles lead to the same  
unique coupling constant. 

The three-point function associated with a $D^*D^*\pi$ vertex  with an
off-shell $D^*$ meson is given by
\begin{equation}
\Gamma^{({D^*})}_{\mu\nu}(p,\pli)=\int d^4x \, d^4y \, \langle 0|T\{j_5(x)
j_{\nu}(y)j^\dagger_\mu(0)\}|0\rangle  
\, e^{ip^\prime.x} \, e^{i(p-\pli).y}\; , 
\label{cor}
\end{equation}
where $j_5=i\bar{u}\gamma_5 d$, $j_\nu=\bar{c}\gamma_\nu u $ and
$ j_\mu=\bar{c}\gamma_\mu d$ are the interpolating fields for the $\pi^-$, 
$D^{*0}$, and $D^{*-}$ respectively with $u$, $d$ and $c$ being the up, down, 
and charm quark fields.

The phenomenological side of the vertex function, $\Gamma_{\mu\nu}
(p,p^\prime)$,
is obtained by the consideration of $\pi$ and $D^*$ state contributions to
the matrix element in Eq.~(\ref{cor}):

\beqa
\Gamma_{\mu\nu}^{(phen)}(p,\pli)&=&{m_\pi^2 m_{D^*}^2\over m_u+m_d}{F_\pi f_{D^*}^2 
g^{(D^*)}_{D^*D^*\pi}(q^2)\over (q^2-m_{D^*}^2)
(p^2-m_{D^*}^2)({\pli}^2-m_\pi^2)}\times
\nonumber \\*[7.2pt]
&&\varepsilon^{\mu\nu\lambda\delta}p_\lambda {\pli}_\delta
+ \mbox{higher resonances}\; .
\label{phen}
\eeqa

To derive Eq.~(\ref{phen}) we have made use of
\beq
\langle D^{*-}(p)|\pi^-(p') D^{*0}(q)\rangle = 
ig^{(D^*)}_{D^* D^*  \pi}(q^2)\varepsilon^{\alpha\gamma\lambda\delta}p_\delta q_\lambda
\epsilon_\alpha(p)\epsilon_\gamma^*(q)
\eeq
where $q=p-\pli$, and the decay constants $F_\pi$ and $f_{D^*}$ are defined by 
the matrix elements
\beq
\langle 0|j_5|\pi(\pli)\rangle={m_\pi^2 F_\pi\over m_u+m_d},
\label{fpi}
\eeq
and
\beq
\langle 0|j_\mu|D^*(p)\rangle=m_{D^*}f_{D^*}\epsilon_\mu(p)
\; ,
\label{fd*}
\eeq
where $\epsilon^\nu$ is the polarization of the vector meson.
The contribution of higher resonances and continuum in Eq.~(\ref{phen})
will be taken into account as usual in the standard form of 
ref.~\cite{io2}, through the continuun thresholds $s_0$ and $u_0$, for the 
$D^*$ and $\pi$ mesons respectively.

The QCD side, or theoretical side, of the vertex function is evaluated by
performing Wilson's operator product expansion (OPE) of the operator
in Eq.~(\ref{cor}). Writing $\Gamma_{\mu\nu}$ in terms of the invariant
amplitude,
\beq
\Gamma_{\mu\nu}(p,\pli)=\Lambda(p^2,{\pli}^2,q^2)
\varepsilon^{\mu\nu\lambda\delta}p_\lambda \pli_\delta,
\eeq
we can write a double dispersion relation for  the invariant
amplitude, $\Lambda$, over the virtualities $p^2$ and ${\pli}^2$
holding $Q^2=-q^2$ fixed:
\beq
\Lambda^{(D^*)}(p^2,{\pli}^2,Q^2)=-{1\over4\pi^2}\int_{m_Q^2}^{s_0} ds
\int_0^{u_0} du {\rho(s,u,Q^2)\over(s-p^2)(u-{\pli}^2)}\;,
\label{dis}
\eeq
where $\rho(s,u,Q^2)$ equals the double discontinuity of the amplitude
$\Gamma(p^2,{\pli}^2,Q^2)$ on the cuts $m_Q^2\leq s\leq\infty$,
$0\leq u\leq\infty$, which can be evaluated using Cutkosky's rules.
Finally, in order to suppress the condensates of higher dimension and at the 
same time reduce the influence of higher resonances  we perform a  
standard Borel transform \cite{svz}: 
\beq
\Pi (M^2) \equiv \lim_{n,Q^2 \rightarrow \infty} \frac{1}{n!} (Q^2)^{n+1} 
\left( - \frac{d}{d Q^2} \right)^n \Pi (Q^2)
\label{borel}
\eeq
($Q^2 = - q^2$ and the squared Borel mass scale $M^2 = Q^2/n$ is kept 
fixed)  in both variables
$P^2=-p^2\rightarrow M^2$ and ${P^\prime}^2=-{\pli}^2\rightarrow \mli$ and 
equate the two representations
described above. We get the following sum rule:
\beqa
& & \frac{m^2_{\pi} m^2_{D^*}}{m_u+m_d} F_{\pi} f^2_{D^*} g^{(D^*)}_{\pi D^* D^*}(Q^2) 
e^{ -m^2_{\pi}/M^{'2} } e^{ - m^2_{D^*}/M^2 }  =  
(Q^2 + m^2_{D^*})  \,\,  \bigg[ <\bar{q}q> \exp(-m^2_c/M^{2}) \nonumber \\
& & - \frac{1}{4 \pi^2} \int_{m^2_c}^{s_0} 
ds  \int_0^{u_{max}} \, 
d u \exp(-s/M^2) \exp(-u/M^{'2}) 
f(s,t,u) \theta(u_0 -u) \bigg]
\label{dsoff}
\eeqa
where $t=-Q^2$, 
\beq
f(s,t,u)={3m_c u(2m_c^2-s-t+u)\over[\lambda(s,u,t)]^{3/2}},
\eeq
$\lambda(s,u,t)=s^2+u^2+t^2-2su-2st-2tu$, and $u_{max}=s+t-m_c^2-{st\over 
m_c^2}$.

We use  the standard values for the numerical parameters:  
$m_{D^*}=2.01\,\GeV$, $m_\pi=140\,\MeV$, $F_\pi=\sqrt{2}\times93\,\MeV$,
$f_{D^{*}}=240~\MeV $, $m_u+m_d=14~\MeV$,
$m_c=1.3\,\GeV$, $\langle\overline{q}q\rangle\,=\,-(0.23)^3\,\GeV^3$.
For the continuum thresholds we take $s_0=(m_{D^*}+\Delta_s)^2$ with 
$\Delta_s=0.5\pm0.1~\GeV$ and $u_0=1.4\pm0.2\,\GeV^2$.

In Fig.~4 we show the perturbative (dotted line) and 
the quark condensate (dashed line) contributions to the form factor 
$g^{(D^{*})}_{\pi D^{*} D^{*}}(Q^2)$ at $Q^2=0.5~\GeV^2$ as a function of the
Borel mass $M^2$ at a fixed ratio ${M^\prime}^2/M^2=0.64/(m_{D^*}^2-m_c^2)$.
We see that the quark condensate contribution is bigger than the perturbative
contribution for values of the Borel mass smaller than $\sim4.5~\GeV^2$.
However, the sum of the both contributions for the form factor, is a very 
stable result as a function of the Borel mass. The quark condensate 
contribution grows with $Q^2$ while the perturbative contribution decreases.
This imposes a limitation over
the region of $Q^2$ that we can use to study the $Q^2$ dependence
of the form factor. Fixing $M^2=10~\GeV^2$, in Fig.~5 we show, through the
filled circles, the momentum dependence of $g^{(D^{*})}_{\pi D^{*} D^{*}}(Q^2)$.

Since the present approach can not be used at $Q^2<0$, in order to extract
the $g_{\pi D^{*} D^{*}}$ coupling from the form factor,  we need to extrapole
the curve to the mass of the off-shell meson $D^*$. In order to do this we fit
 the QCDSR results with an analitycal expression. We have obtained a good fit
using a exponential form:
\beqa
g^{(D^{*})}_{\pi D^{*} D^{*}}(Q^2)=4.8 ~e^{-Q^2 / 6.8}~\GeV^{-1}\;,
\label{dsdsof}
\eeqa
where $6.8$ is in units of $\GeV^2$
This fit is also shown in Fig.~5 through the solid line. From Eq.(\ref{dsdsof})
we get 
$g_{\pi D^{*} D^{*}}=g^{(D^{*})}_{\pi D^{*} D^{*}}(Q^2=-m_{D^*}^2)=8.7~
\GeV^{-1}$. To check the consistency of the calculation, we also evaluate the
form factor at the same vertex, but for an off-shell pion. In this case we have
to evaluate the three-point function
\begin{equation}
\Gamma^{(\pi)}_{\mu\nu}(p,\pli)=\int d^4x \, d^4y \, \langle 0|T\{j_{\nu}(x)
j_5(y)j^\dagger_\mu(0)\}|0\rangle  
\, e^{ip^\prime.x} \, e^{i(p-\pli).y}\; . 
\label{corpi}
\end{equation}

Proceeding in a similar way we obtain the following sum rule:
\beqa
%& & 
\frac{m^2_{\pi} m^2_{D^*}}{m_u+m_d} F_{\pi} f^2_{D^*} g^{(\pi)}_{\pi D^* 
D^*}(Q^2) 
e^{ -m^2_{D^*}/M^{'2} } e^{ - m^2_{D^*}/M^2 }  =  
(Q^2 + m^2_{\pi})  \,\,  \bigg[ \nonumber \\
%& & 
- \frac{1}{4 \pi^2} \int_{s_{min}}^{s_0} 
ds  \int_{u_{min}}^{u_0} \, 
d u~ e^{-s/M^2} e^{-u/M^{'2}} {3m_c t(s+u-t-2m_c^2)\over[\lambda(s,u,t)]^{3/2}}
%g(s,t,u)  
\bigg]
\label{pioff}
\eeqa
where  
$u_{min}=m_c^2-{m_c^2t\over s- m_c^2}$ and $s_{min}=m_c^2(1-{t\over u_0- m_c^2}
\bigg)$. Now we use $u_0=(m_{D^*}+\Delta_u)^2$ with 
$\Delta_u=0.5\pm0.1~\GeV$, and $M^2=M^{'2}$. The results are also rather stable
as a function of the Borel mass. 
% and, fixing $M^2=7~\GeV^2$, we show in Fig.~5,
%through the squares, the QCDSR results for $g^{(\pi)}_{\pi D^* D^*}(Q^2)$. 
We also got a good fit of the QCDSR results for $g^{(\pi)}_{\pi D^* D^*}(Q^2)$
using a exponential form:
\beqa
g^{(\pi)}_{\pi D^{*} D^{*}}(Q^2)=8.5 ~e^{-Q^2 / 3.4}~\GeV^{-1},
\label{dsdspiof}
\eeqa
where $3.4$ is in units of $\GeV^2$. This fit is also shown in Fig.~5 through 
the dot-dashed line. From Eq.(\ref{dsdspiof})
we get 
$g_{\pi D^{*} D^{*}}=g^{(\pi)}_{\pi D^{*} D^{*}}(Q^2=-m_{\pi}^2)=8.5~
\GeV^{-1}$.

Considering the uncertainties in the continuum thresholds, and the difference
in the values of the coupling extracted when the $D^*$ or $\pi$ mesons
are off-shell, our result for the $\pi D^* D^*$ coupling constant is
\beqa
g_{\pi D^{*} D^{*}}=8.6 \pm 1.0~\GeV^{-1},
\label{dsdspiconst}
\eeqa

The triple vertex couplings were 
calculated as explained above. The quartic vertex couplings could not be obtained with QCDSR 
and we  have used the prescription given in \cite{osl}:
\beqa
g_{\psi D D \pi}=\biggr ( \frac{\sqrt{3}}{6}-\frac{1}{4}\biggr)\,
\frac{g_a\,N_c}{16\,\pi^2\,F_{\pi}^3}
\eeqa
\beqa
g_{\psi D^* D \pi}=\frac{1}{2}\,g_{\psi D D}\,g_{D^* D \pi}
\eeqa
\beqa
g_{\psi D^* D^* \pi}=\frac{1}{2}\,\frac{g_a^3\, N_c}{32\,\pi^2\,F_{\pi}}
\eeqa
where $g_a$ is obtained from:
\beqa
g_{\psi D^* D}=\frac{\sqrt{2}}{4 \sqrt{3}}\frac{g_a^2\,N_c}{16\,\pi^2\,F_{\pi}}
\eeqa
In the above expressions  $N_c=3$ and the triple vertex couplings are taken from our 
calculations. For completeness we present in Table I all the couplings. 

\begin{table}[htb]
%\begin{center}
\begin{tabular}{|c|c|} \hline
$g_{D^*D\pi}$ & $17.9$ \\
\hline
$g_{\psi D^{*}D}$ & $4.0$ $GeV^{-1}$ \\
\hline
$g_{\psi DD}$ & $5.8$  \\
\hline
$g_{\psi D^*D^*}$ & $6.2$ \\
\hline
$g_{D^*D^* \pi}$ & $8.6$ $GeV^{-1}$ \\
\hline
\hline
$g_{\psi D D \pi}$ & $10.0$ $GeV^{-3}$ \\
\hline
$g_{\psi D^* D \pi}$ & $51.9$ \\
\hline
$g_{\psi D^* D^* \pi}$ & $57.0$ $GeV^{-1}$ \\
\hline
\end{tabular}
%\end{center}
\caption{Coupling constants used in the numerical calculations.  The first four
were calculated with QCDSR and the last three were obtained with the prescription of \protect\cite{osl}}
\end{table}

\section{The cross sections}

Having all the needed form factors, we now proceed to the  evaluation of the cross 
sections. As in previous calculations, these cross sections for secondary $J/\psi$ 
production will be related to  the annihilation through detailed balance:
\beqa
\sigma_{(3+4\rightarrow 1+2)}=\sigma_{(1+2\rightarrow 3+4)}\,
\frac{(2S_1+1)(2S_2+1)}{(2S_3+1)(2S_4+1)}\,\frac{P_i^2}{P_f^2}
\label{bal}
\eeqa

In figure~6 we show the $J/\psi$ secondary production cross section as a 
function of $\sqrt{s}$, without form factors. In all figures, the channels 
$D \bar{D}\rightarrow J/\psi+\pi$, $D \bar{D^*}\rightarrow J/\psi+\pi$ and 
$D^* \bar{D^*}\rightarrow J/\psi+\pi$ are represented by solid, dashed and dotted 
lines respectively. In figure 7, with the help of (\ref{bal}) we show the 
corresponding inverse reactions. As it can be seen, the cross sections have the 
same order of magnitude in both directions. Figures 8 and 9 are the analogues of 
6 and 7 when we include the form factors in the calculations.  Of course, as we 
stressed in the introduction, only these last two figures correspond to realistic 
numbers. The comparison of the two sets of figures is interesting only to estimate 
the effect of form factors. In previous studies doing the same kind of comparison, as
for example in \cite{osl}, the introduction of form factors reduced the cross 
sections 
by factors ranging between  $20$  and  $50$ depending on the channel. In that work 
the form factor was the same for all 
vertices and the cut-off,  not known,   was estimated to be between $1$ and $2$ GeV. 
Our study is much more detailed and not only each vertex has its own form factor,
but, depending on which particle is off-shell the form factor is different. The final
effect of all these peculiarities is the reduction of the cross sections by a factor
around $7$. Although significant, this reduction is smaller than previously expected.

Figure 8 contains our main results. The plotted cross sections can be compared with 
the results of \cite{ko98} and, more directly,  with \cite{brat}. In figure 2 of 
\cite{ko98}, although the variables in the plot are different, we can observe the same 
trend and relative importance of the three channels. In that work, the results were obtained 
with the quark model of  \cite{mbq}. Our curves share some features with the results of 
\cite{brat}, such as, for example, the dominance of the $D D^*$ channel and the falling 
trend of the $DD^*$ and $D^* D^*$ channels.  The behavior of the $DD$ channel is quite 
different. In the energy range of $\sqrt{s} > 4.5$ GeV our cross sections are smaller by a
factor of $2$ ($DD^*$) or $5$ ($D^* D^*$ and $D D$). These discrepancies are large but they 
are expected  since in \cite{brat} all channels include the final state $J/\psi + \rho$. We 
could not include it consistently because the form factors of the $\rho D D^*$ and 
$\rho D^*  D^*$ vertices have not yet been studied with our techniques and are thus not
available. In the model used by the Giessen group the cross sections for 
$D + \bar{D}  \rightarrow J/\psi + \pi$ and $D + \bar{D} \rightarrow  J/\psi + \rho$ are 
similar and the same conclusion holds for the other inital state open charm mesons.  If
this would remain true  in the effective lagrangian approach, then our results 
including both final states would come closer to those of \cite{brat}, giving thus a more 
theoretical support to the model considered there.

\section{Summary and conclusions}

We have updated the calculations of the cross sections for $J/\psi$ dissociation and 
production in the effective lagrangian approach.  The novel feature introduced in this 
work is the use of the  form factors (\ref{h1}) - (\ref{h6}) and especially 
(\ref{dsdsof}), which was obtained here from QCDSR. We believe that our results are useful 
for numerical simulations of heavy ion collisions, such as those performed in 
\cite{thews,rapp,polleri,brat} and also in \cite{ko98,redlich,zhang2002}. The calculation 
of the cross sections of the  processes considered here  are complete. However, our program 
is not yet finished and there are still form factors to be calculated, such as  $\rho D D^*$ 
and $\rho D^*  D^*$. These calculations are under way and we will eventually have all 
hadronic form factors.

Although no strong statement can be made without knowing the 
$D + \bar{D} \rightarrow  J/\psi + \rho$ cross section, our results give partial support to
the conclusion advanced in \cite{brat}, namely that the open charm fusion cross sections are 
large enough to produce a sizeable number of ``recreated''  $J/\psi$'s already in heavy ion
collisions at RHIC.

\underline{Acknowledgements}: 
This work has been supported by CNPq and  FAPESP. We are deeply grateful to 
R. Azevedo for fruitful discussions and help with numerical codes. 

%\end{thebibliography}

%\eject

\newpage

\begin{figure} \label{fig1}
%\leavevmode
\centerline{\psfig{figure=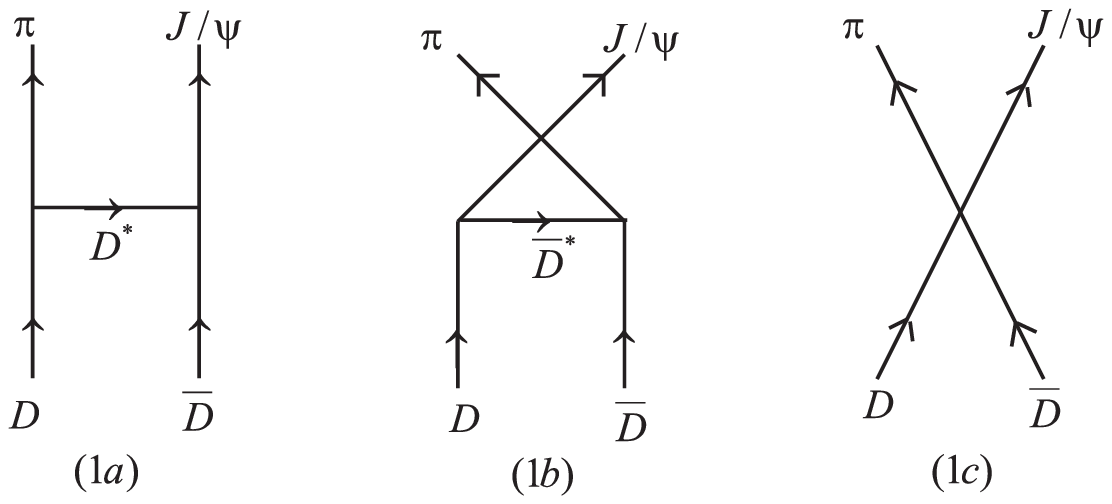,width=14cm,angle=0}}
\caption{Diagrams which contribute to the process 
$D \bar{D}\rightarrow J/\psi+\pi$}
\end{figure}

\begin{figure} \label{fig2}
%\leavevmode
\centerline{\psfig{figure=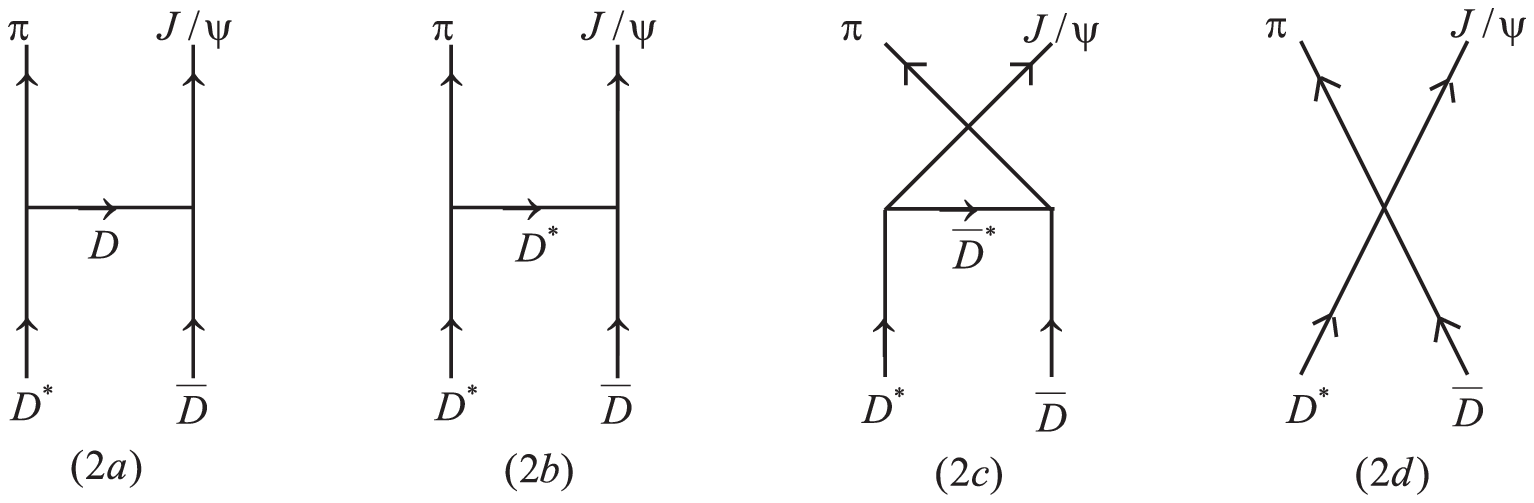,width=14cm,angle=0}}
\caption{Diagrams which contribute to the process
$D^* \bar{D}\,\rightarrow J/\psi + \pi$}
\end{figure}

\begin{figure} \label{fig3}
%\leavevmode
\centerline{\psfig{figure=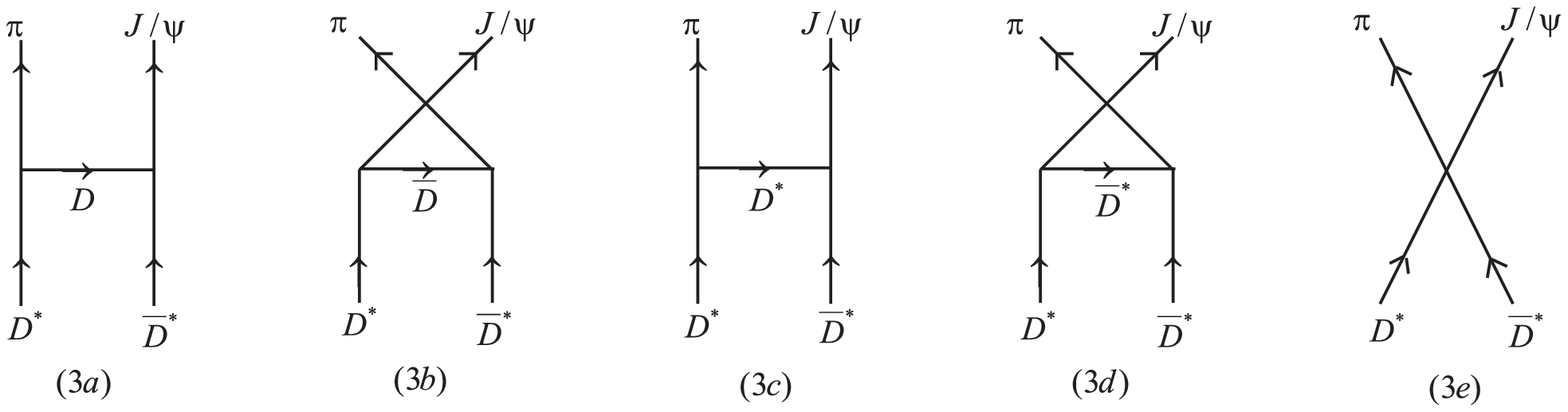,width=14cm,angle=0}}
\caption{ Diagrams which contribute to the process
$D^* \bar{D^*}\,\rightarrow J/\psi + \pi$}
\end{figure}

\begin{figure} \label{fig4}
%\leavevmode
\centerline{\psfig{figure=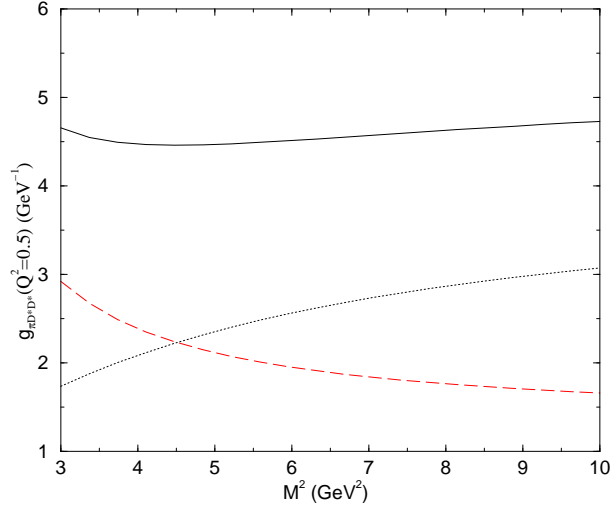,width=8cm,angle=0}}
\caption{$M^2$ dependence of the perturbative contribution (dotted line) and 
the quark condensate contribution (dashed line) to the 
$g^{(D^{*})}_{\pi D^{*} D^{*}}(Q^2)$ at $Q^2=0.5~\GeV^2$. The solid line 
gives the final result for the form factor.}
\end{figure}

\begin{figure} \label{fig5}
%\leavevmode
\centerline{\psfig{figure=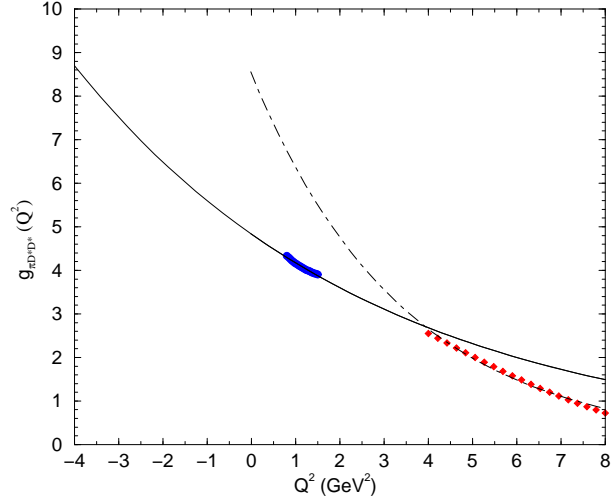,width=8cm,angle=0}}
\caption{Momentum dependence of the $\pi D^{*} D^{*}$ form factors. The
solid and dot-dashed lines give the parametrization of the QCDSR results for
$g^{(D^{*})}_{\pi D^{*} D^{*}}(Q^2)$ (circles) and $g^{(\pi)}_{\pi D^{*} 
D^{*}}(Q^2)$ (squares) respectively.}
\end{figure}

\begin{figure} \label{fig6}
%\leavevmode
\centerline{\psfig{figure=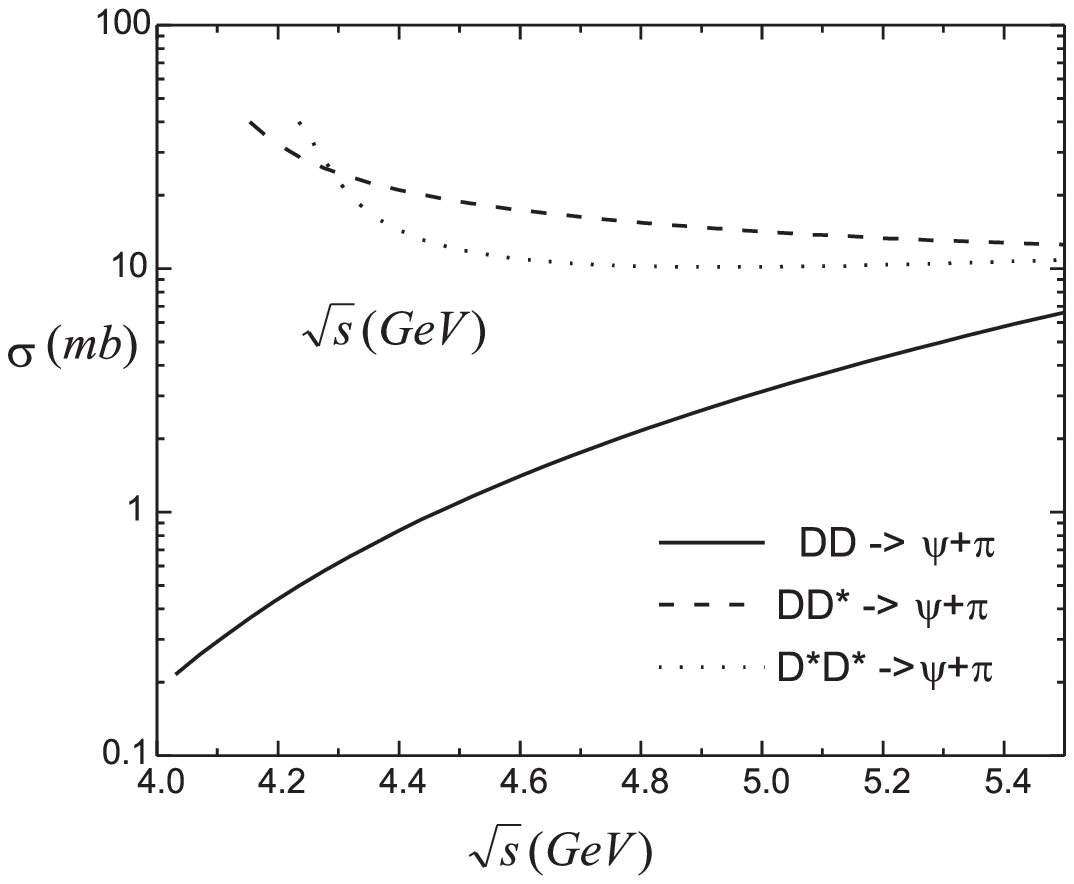,width=8cm,angle=0}}
\caption{$J/\psi$ secondary production cross section without form factors}
\end{figure}

\begin{figure} \label{fig7}
%\leavevmode
\centerline{\psfig{figure=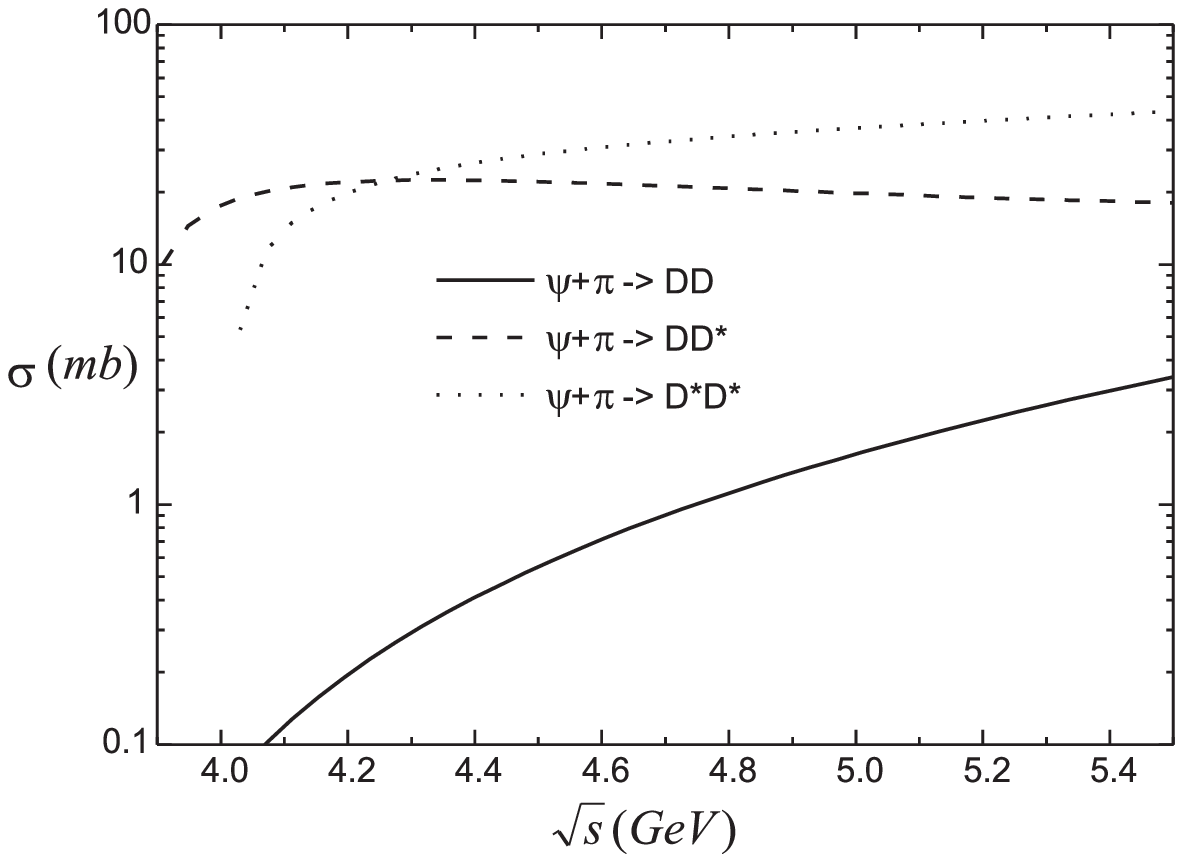,width=8cm,angle=0}}
\caption{$J/\psi$ absorption cross section obtained through detailed balance without
form factors}
\end{figure}

\begin{figure} \label{fig8}
%\leavevmode
\centerline{\psfig{figure=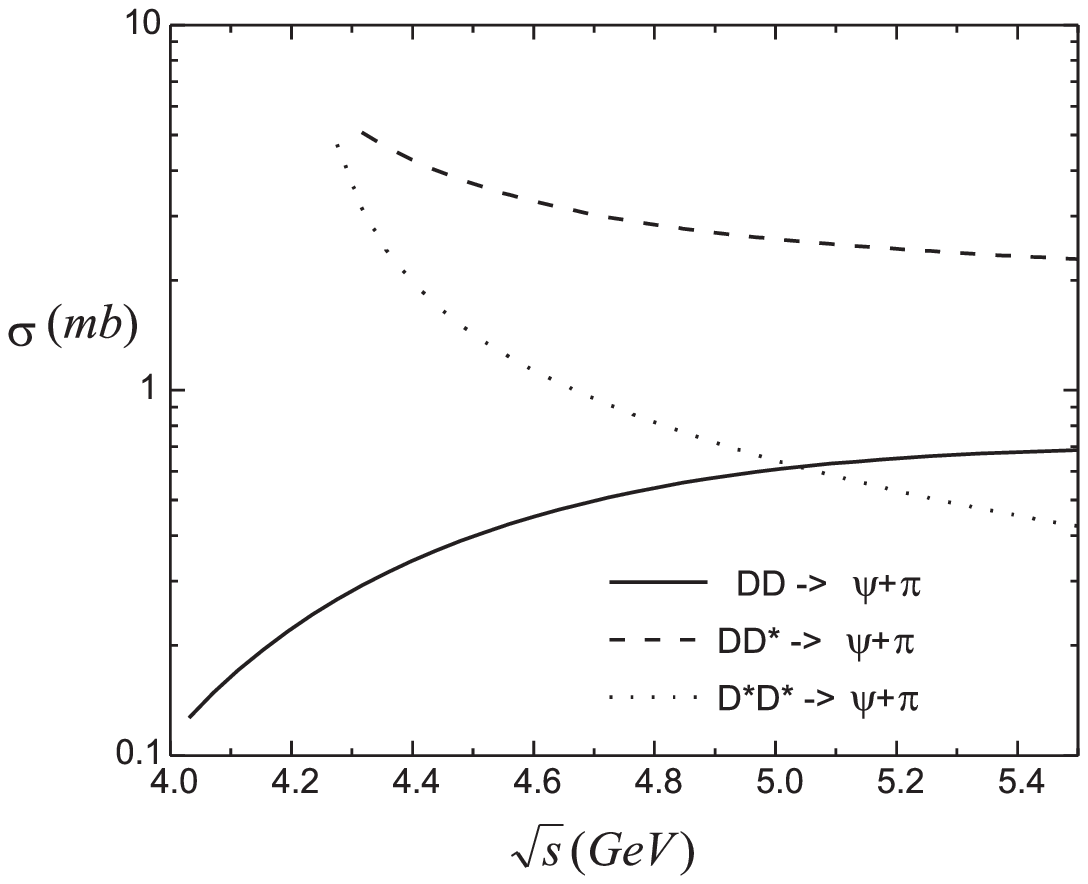,width=8cm,angle=0}}
\caption{$J/\psi$ secondary production cross section with form factors}
\end{figure}

\begin{figure} \label{fig9}
%\leavevmode
\centerline{\psfig{figure=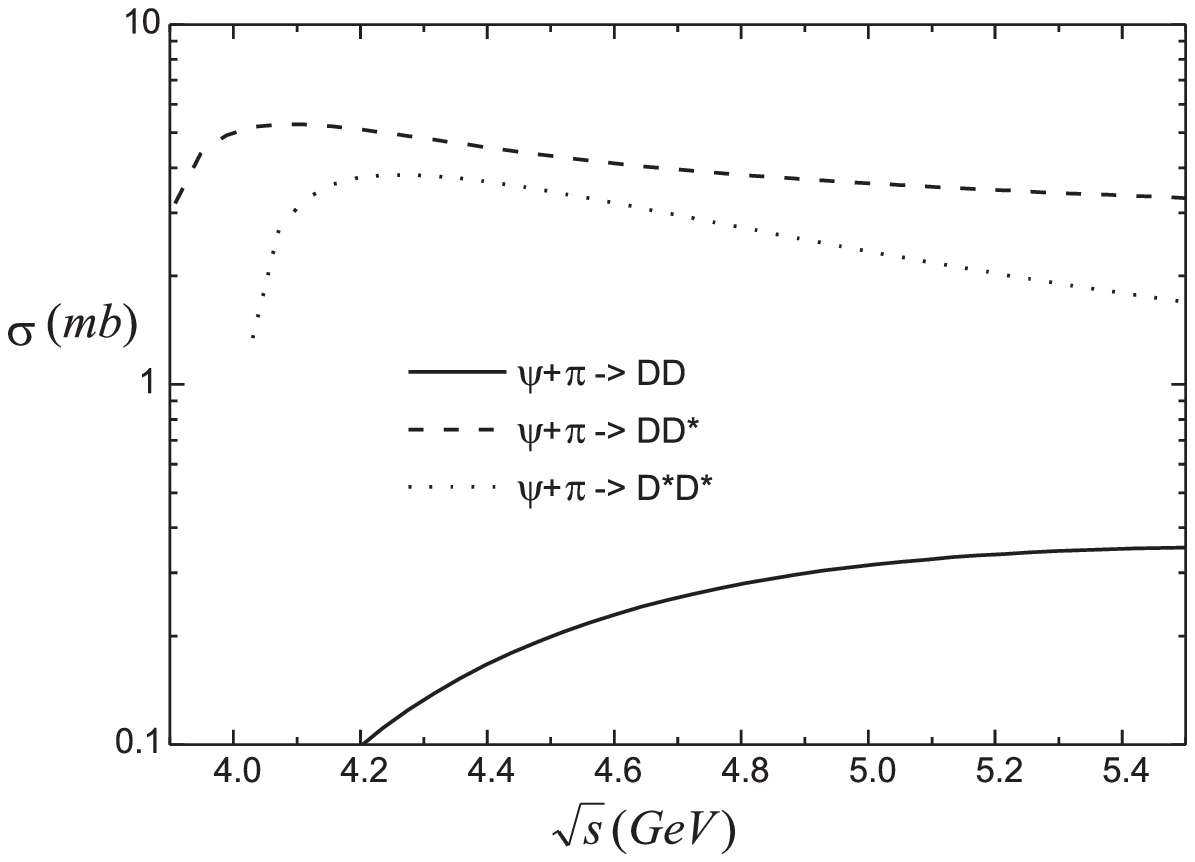,width=8cm,angle=0}}
\caption{$J/\psi$ absorption cross section obtained through detailed balance with
form factors}
\end{figure}

\end{document}